\documentclass{article}

\usepackage{amsmath,amssymb}
\usepackage{graphicx}
\usepackage{color}
\usepackage[all]{xy}

\newtheorem{theorem}{Theorem}

\newtheorem{lemma}{Lemma}

\newtheorem{definition}{Definition}
\newtheorem{proposition}{Proposition}
\newtheorem{remark}{Remark}

\newtheorem{condition}{Condition}
\newtheorem{openproblem}{Open Problem}

\newcommand{\done}{\hfill $\Box$ }

\newcommand{\ls}[1]
    {\dimen0=\fontdimen6\the\font\lineskip=#1\dimen0
     \advance\lineskip.5\fontdimen5\the\font
     \advance\lineskip-\dimen0
     \lineskiplimit=0.9\lineskip
     \baselineskip=\lineskip
     \advance\baselineskip\dimen0
     \normallineskip\lineskip\normallineskiplimit\lineskiplimit
     \normalbaselineskip\baselineskip
     \ignorespaces}

\begin{document}

\bibliographystyle{abbrv}

\title{Two Problems about Monomial Bent Functions}

\author{Honggang Hu$^1$, Bei Wang$^1$, Xianhong Xie$^1$, and Yiyuan Luo$^2$
\smallskip\\
$^1$School of Information Science and Technology\\
University of Science and Technology of China\\
Hefei, China, 230027\\
Email. hghu2005@ustc.edu.cn, \{wangbei,xianhxie\}@mail.ustc.edu.cn
\smallskip\\
$^2$School of Computer Science and Engineering\\
Huizhou University\\
Huizhou, China, 516007\\
Email. luoyy@hzu.edu.cn}

\date{}
 \maketitle

\thispagestyle{plain}
\setcounter{page}{1}

\begin{abstract}
In 2008, Langevin and Leander determined the dual function of three classes of monomial bent functions with the help of Stickelberger's theorem:
Dillon, Gold and Kasami. In their paper, they proposed one very strong condition such that their method works, and showed that both Gold exponent and Kasami exponent satisfy this condition. In 2018, Pott {\em et al.} investigated the issue of vectorial functions with maximal number of bent components. They found one class of binomial functions which attains the upper bound. They also proposed an open problem regarding monomial function with maximal number of bent components.

In this paper, we obtain an interesting result about the condition of Langevin and Leander, and solve the open problem of Pott {\em et al.}. Specifically, we show that: 1) for a monomial bent function over $\mathbb{F}_{2^{2k}}$, if the exponent satisfies the first part of the condition of Langevin and Leander, then it satisfies the entire condition; 2) $x^{2^k+1}$ is the only monomial function over $\mathbb{F}_{2^{2k}}$ which has maximal number of bent components. Fortunately, as a consequence, we also solve an open problem of Ness and Helleseth in 2006.
\end{abstract}

{\bf Key Words. }Bent function, finite field, Hamming weight, Walsh transform, vectorial function.

\ls{1.5}
%===========================================================================
%===========================================================================
\section{Introduction}\label{sec_intro}

Bent functions have several areas of application such as cryptography, coding theory and communications because they have the maximum nonlinearity \cite{GG05,HK98}. The nonlinearity measures the distance of a Boolean function to the set of affine functions. Since Dillon and Rothaus introduced bent functions firstly \cite{Dillon72,Rothaus76}, this special class of Boolean functions has been an interesting research issue for more than 40 years \cite{CCK08,CG06,CG08,DLCCFG06,LL08,Leander06,LK06,Mesnager11,MF13,MPB14,PPMB18,YG01}. For more history on bent functions, the reader is referred to \cite{Mesnager16}. There are also some works about generalized bent functions over finite fields \cite{HeHoKhWaXi09,HeKh06,HeKh10,HYT18,HZS17}.

Let $\mathbb{F}_{2^n}$ be the finite field of $2^n$ elements, and $\mathbb{F}_{2^n}^{*}$ be the set of nonzero elements in $\mathbb{F}_{2^n}$. If we fix one basis of $\mathbb{F}_{2^n}$ over $\mathbb{F}_2$, then any Boolean function from $\mathbb{F}_2^n$ to $\mathbb{F}_2$ can be viewed as a function from $\mathbb{F}_{2^n}$ to $\mathbb{F}_2$. Let $Tr_1^n(\cdot)$ be the trace function from $\mathbb{F}_{2^n}$ to $\mathbb{F}_2$, and $f(x)$ be a function from $\mathbb{F}_{2^n}$ to $\mathbb{F}_2$. Then the Hadamard transform of $f(x)$ is defined by
$$\widehat{f}(\lambda)=\sum_{x\in\mathbb{F}_{2^n}}(-1)^{f(x)+Tr_1^n(\lambda x)},$$
where $\lambda\in \mathbb{F}_{2^n}$. $f(x)$ is a bent function if and only if $\widehat{f}(\lambda)=\pm 2^{n/2}$ for any $\lambda\in \mathbb{F}_{2^n}$.

A function $f(x)$ from $\mathbb{F}_{2^n}$ to $\mathbb{F}_2$ is called a monomial function if $f(x)=Tr_1^n(\alpha x^d)$, where $\alpha\in \mathbb{F}_{2^n}^{*}$ and $0<d<2^n-1$. Among all kinds of bent functions, monomial bent functions are especially interesting. For any integer $r$, let $(\mathbb{F}_{2^n})^r$ denote the set $\{ y^r | y\in \mathbb{F}_{2^n}\}$. So far, there are five classes of known monomial bent functions over $\mathbb{F}_{2^n}$ with $n=2k$, and they are listed in \cite{MP13}:
\begin{itemize}
  \item[1)] $d=2^t+1$ with $n/\gcd(n, t)$ being even and $\alpha\not\in (\mathbb{F}_{2^n})^d$ \cite{Gold68};
  \item[2)] $d=r(2^k-1)$ with $\gcd(r, 2^k+1)=1$ and $\alpha\in \mathbb{F}_{2^k}^{*}$ satisfying $\sum_{x\in\mathbb{F}_{2^k}^{*}}(-1)^{Tr_1^k(\alpha x+1/x)}=-1$ \cite{CG08,Dillon74};
  \item[3)] $d=2^{2t}-2^t+1$ with $\gcd(t, n)=1$ and $\alpha\not\in (\mathbb{F}_{2^n})^3$ \cite{DD04,LL08};
  \item[4)] $d=(2^t+1)^2$ with $t$ odd, $n=4t$, and $\alpha\in \gamma \mathbb{F}_{2^t}$ satisfying $\gamma\in \mathbb{F}_{2^2}\setminus \mathbb{F}_{2}$ \cite{CK08,Leander06};
  \item[5)] $d=2^{2t}+2^t+1$ with $t>1$, $n=6t$, and $\alpha\in \mathbb{F}_{2^{3t}}^{*}$ satisfying $Tr_t^{3t}(\alpha)=0$ \cite{CCK08}.
\end{itemize}

In 2008, Langevin and Leander investigated the dual function of monomial bent functions, and solved three cases: Dillon, Gold and Kasami \cite{LL08}. With the help of Stickelberger's theorem and Teichm\"{u}ller character, they found a general way to compute the dual function of monomial bent functions over finite fields with characteristic 2. They need a very strong condition such that their approach works. Fortunately, they proved that both Gold exponent and Kasami exponent satisfy this condition. In particular, for Kasami exponent, via this approach Langevin and Leander obtained a more general result than that of Dillon and Dobbertin in \cite{DD04}.

For a vectorial function from $\mathbb{F}_2^n$ to $\mathbb{F}_2^m$, if all nonzero linear combinations of the coordinate functions of this function are also bent, it is called a vectorial bent function. Similar to the case of Boolean functions, any vectorial function from $\mathbb{F}_2^n$ to $\mathbb{F}_2^m$ can be viewed as a function from $\mathbb{F}_{2^n}$ to $\mathbb{F}_{2^m}$. Let $F(x)$ be a vectorial function from $\mathbb{F}_{2^n}$ to $\mathbb{F}_{2^m}$. Then $F(x)$ is a vectorial bent function if and only if $Tr_1^m(\alpha F(x))$ is a bent function for any $\alpha\in \mathbb{F}_{2^m}^{*}$. Nyberg showed that such vectorial bent functions can only exist for the case of $n\geq 2m$ \cite{Nyberg91}. Two different constructions of vectorial bent functions from some known classes of bent functions have also been proposed in \cite{Nyberg91}. In 2014, vectorial bent functions from multiple terms trace functions have been studied \cite{MPB14}. In 2018, Pott {\em et al.} investigated vectorial functions from $\mathbb{F}_{2^n}$ to $\mathbb{F}_{2^n}$ with maximal number of bent components \cite{PPMB18}. They proved that the number of bent components of any vectorial function is at most $2^{2k}-2^k$, where $n=2k$. They found one class of binomial functions which attains this upper bound. Furthermore, they also presented an open problem to show that $x^{2^k+1}$ is the only monomial function over $\mathbb{F}_{2^{2k}}$ with maximal number of bent components.

In this paper, we investigate the strong condition of Langevin and Leander in 2008, and solve the open problem of Pott {\em et al.} in 2018 \cite{LL08,PPMB18}. Let $n=2k$. Our main contributions are as follows.
\begin{enumerate}
  \item[1)] We dig out an interesting result about the condition of Langevin and Leander. Suppose that $d$ is the exponent of a monomial bent function over $\mathbb{F}_{2^n}$. For any integer $j$, let $\mathrm{wt}(j)$ be the Hamming weight of $j$ modulo $2^n-1$ which will be defined later. If $\min_{0<j<2^n-1}\mathrm{wt}(j)+\mathrm{wt}(-jd)=k$, and $\mathrm{wt}(\widetilde{j})+\mathrm{wt}(-\widetilde{j}d)=k$ with $0<\widetilde{j}<2^n-1$, then we show that $\widetilde{j}d\equiv0\mod(2^n-1)$. This result means that if $d$ satisfies the first part of the condition, then it satisfies the entire condition.
  \item[2)] For the open problem of Pott {\em et al.}, based on one result of Pott {\em et al.} and our new observations, we find nice conditions under which a monomial function over $\mathbb{F}_{2^n}$ has maximal number of bent components. Furthermore, we show that if a monomial function over $\mathbb{F}_{2^n}$ satisfies these conditions, then it must be $x^{2^k+1}$. Because this open problem is essentially the same as an open problem of Ness and Helleseth in 2006 \cite{NH06,ZPKLL20}. Hence, we also solve that open problem.
\end{enumerate}

This paper is organized as follows. In Section \ref{sec_pre}, we provide some necessary notation and background. The strong condition of Langevin and Leander is studied in Section \ref{sec_main_1}, and the open problem of Pott {\em et al.} is solved in Section \ref{sec_main_2}. Finally, Section \ref{sec_con} concludes this paper.

%===========================================================================
%===========================================================================
\section{Preliminaries}\label{sec_pre}

%===========================================================================
\subsection{Cyclotomic Cosets and Trace Representations}

For any $0\leq t<2^n-1$, let $n_t>0$ be the smallest integer such that $t\equiv t2^{n_t}(\mbox{mod }2^n-1)$. Then we know that $n_t|n$. The set $$C_t=\{t, 2t, ..., 2^{n_t-1}t\}$$ is defined to be the cyclotomic coset containing $t$ modulo $2^n-1$, and the
smallest integer in $C_t$ is called the coset leader of $C_t$. For simplicity, we may assume that $t$ is the coset leader of $C_t$.

\begin{proposition}[\cite{GG05}]\label{prop_trace}
Let $f(x)$ be a nonzero function from $\mathbb{F}_{2^n}$ to $\mathbb{F}_2$. Then $f(x)$ can be represented as
$$f(x)=\sum_{j\in \Gamma(n)}Tr_1^{n_j}(F_jx^j)+F_{2^n-1}x^{2^n-1}, F_j\in \mathbb{F}_{2^{n_j}}, F_{2^n-1}\in \mathbb{F}_2$$
where $\Gamma(n)$ is the set of all coset leaders modulo $2^n-1$, $n_j|n$ is the size of the coset $C_j$, and $Tr_1^{n_j}(x)$
is the trace function from $\mathbb{F}_{2^{n_j}}$ to $\mathbb{F}_2$.
\end{proposition}

%===========================================================================
\subsection{Monomial Bent Functions over $\mathbb{F}_{2^{2k}}$}

From now on, let $n=2k$.

For any $0\leq j<2^n-1$, let $j=\sum_{i=0}^{n-1}j_i2^{i}$ be the binary representation of $j$, where $j_i\in\{0, 1\}$ for any $0\leq i<n$. Let $\mathrm{wt}(j)=\sum_{i=0}^{n-1}j_i$. Moreover, for $j<0$ or $j\geq 2^n-1$, we use $\mathrm{wt}(j)$ to denote $\mathrm{wt}(\overline{j})$, where $0\leq \overline{j}<2^n-1$
and $j\equiv \overline{j}\ (\bmod\;2^n-1)$. For $0<d<2^n-2$, let $V_d(j)=\mathrm{wt}(j)+\mathrm{wt}(-jd)$.

In 2008, in order to study monomial bent functions over $\mathbb{F}_{2^{2k}}$, Langevin and Leander proposed the following condition about the exponent $d$.
They proved that both Kasami exponent and Gold exponent satisfy this condition.

\begin{condition}[\cite{LL08}]\label{con}
$\min_{0<j<2^n-1}V_d(j)=k$, and $V_d(\widetilde{j})=k$ with $0<\widetilde{j}<2^n-1\Rightarrow \widetilde{j}d\equiv0\mod(2^n-1)$
\end{condition}

%===========================================================================
\subsection{Vectorial Functions over $\mathbb{F}_{2^{2k}}$ and Bent Components}

For any vectorial function $F(x)$ over $\mathbb{F}_{2^{2k}}$, let $\mathcal{S}_F=\{\ \alpha\ |\ \alpha\in \mathbb{F}_{2^{2k}}, Tr_1^{2k}(\alpha F(x))\mbox{ is not a bent function}\ \}$. In 2018, Pott {\em et al.} proved the following result.

\begin{theorem}[\cite{PPMB18}]\label{thm_S}
With notation as above, we have $|\mathcal{S}_F|\geq 2^k$. In particular, if $|\mathcal{S}_F|=2^k$, then $\mathcal{S}_F$ is a linear space of dimension $k$ over $\mathbb{F}_2$.
\end{theorem}

If $Tr_1^{2k}(\alpha F(x))$ is a bent function, then it is called a bent component of $F(x)$. By Theorem \ref{thm_S}, the number of bent components of $F(x)$ is at most $2^{2k}-2^k$.
In addition to the known case of $x^{2^k+1}$, in \cite{PPMB18}, Pott {\em et al.} showed that $x^{2^i}(x+x^{2^k})$ also have $2^{2k}-2^k$ bent components, where $i\geq0$.

In 2018, Pott {\em et al.} proposed the following open problem.

\begin{openproblem}[\cite{PPMB18}]\label{open}
Show that $x^{2^k+1}$ is the only monomial function over $\mathbb{F}_{2^{2k}}$ which has $2^{2k}-2^k$ bent components.
\end{openproblem}

\begin{remark}
In 2006, Ness and Helleseth studied the cross-correlation function of two binary $m$-sequences with different lengths \cite{NH06}: one length is $2^{2k}-1$, and the other length is $2^k-1$.
They left an open problem: the cross-correlation function takes exactly two different values if and only if the decimation number is $2^i$ for some $i\geq 0$.
In 2020, Zheng et al. pointed out that this open problem is essentially the same as Open Problem \ref{open} \cite{ZPKLL20}. Therefore, if we can solve Open Problem \ref{open}, then we can solve the open problem of Ness and Helleseth in 2006.
\end{remark}

%===========================================================================
\subsection{Gauss Sums and Stickelberger's Theorem}

For a finite field, there are two kinds of characters: additive character and multiplicative character. Let $\psi$ be the mapping defined by
$$\psi(x)=(-1)^{Tr_1^n(x)}.$$
Then $\psi$ is an additive character of $\mathbb{F}_{2^n}$. Let $\chi$ be a multiplicative character of $\mathbb{F}_{2^n}^{*}$.
For simplicity, we define $\chi(0)=0$ which extends $\chi$ to $\mathbb{F}_{2^n}$. For the convenience, we denote the multiplicative character set of $\mathbb{F}_{2^n}^{*}$ by $\widehat{\mathbb{F}_{2^n}^{*}}$.

\begin{definition}[\cite{LN83}]
For any multiplicative character $\chi$ over $\mathbb{F}_{2^n}$, the
Gauss sum $G(\chi)$ over $\mathbb{F}_{2^n}$ is defined by
$$G(\chi)=\sum_{x\in \mathbb{F}_{2^n}}\psi(x)\chi(x).$$
\end{definition}

\begin{lemma}[\cite{LN83}]\label{lem_gauss}
For any multiplicative character $\chi$ over $\mathbb{F}_{2^n}$, we have
$$G(\overline{\chi})=\chi(-1)\overline{G(\chi)}\mbox{ and }G(\chi^2)=G(\chi).$$
If $\chi$ is trivial, then $G(\chi)=-1$. On the other hand, if $\chi$ is
nontrivial, then
$$G(\chi)\overline{G(\chi)}={2^n}.$$
\end{lemma}

Let $\omega_{2^n-1}=e^{2\pi i/(2^n-1)}$, which is the complex primitive $(2^n-1)$-th root of unity.
Then $G(\chi)\in\mathbb{Z}[\omega_{2^n-1}]$ for any multiplicative character $\chi$.
The algebraic integer ring $\mathbb{Z}[\omega_{2^n-1}]$ is very useful for the study of Gauss sums over $\mathbb{F}_{2^n}$. In $\mathbb{Z}$, $(2)$ is a prime ideal. Let $t=\phi(2^n-1)/n$. Then $(2)$ can be factored into the product of $t$ different prime ideals in $\mathbb{Z}[\omega_{2^n-1}]$, i.e., $(2)=\mathcal{P}_1\mathcal{P}_2\cdots  \mathcal{P}_t$, where $\mathcal{P}_i$ is a prime ideal in $\mathbb{Z}[\omega_{2^n-1}]$ for any $1\leq i\leq t$. For each $\mathcal{P}_i$, it holds that $$\mathbb{Z}[\omega_{2^n-1}]/\mathcal{P}_i\cong \mathbb{F}_{2^n},$$
because $[\mathbb{Z}[\omega_{2^n-1}]/\mathcal{P}_i:\mathbb{Z}/(2)]=n$. Henceforth, we fix one prime ideal
$\mathcal{P}_i$ and denote it by $\mathcal{P}$ for simplicity.

There is one special multiplicative character $\chi$
on $\mathbb{F}_{2^n}$ satisfying
$$\chi(x)(\mbox{mod }\mathcal{P})=x.$$
This character is called the Teichm\"{u}ller character, and we denote it by $\chi_\mathfrak{p}$. The Teichm\"{u}ller character can generate the group $\widehat{\mathbb{F}_{2^n}^{*}}$. According to Stickelberger's theorem \cite{Lang78}, for any $0<j<2^n-1$, we have
$$G(\chi_\mathfrak{p}^{-j})\equiv 2^{\mathrm{wt}(j)}\mbox{\ mod }2^{\mathrm{wt}(j)+1}.$$
Stickelberger's theorem is very important for the proofs of three nice conjectures: Welch, Niho, and Lin conjectures \cite{Mc72,CCD00,HoXi01,HSGH14}.

The lemma below is pretty useful and well known.
\begin{lemma}\label{lem_gauss_trace}
For any $x\in \mathbb{F}_{2^n}^{*}$, we have
$$(-1)^{Tr_1^n(x)}=\frac{1}{2^n-1}\sum_{\chi\in\widehat{\mathbb{F}_{2^n}^{*}}}G(\chi)\overline{\chi}(x).$$
\end{lemma}

%===========================================================================
\subsection{The Binary Modular Add-With-Carry Algorithm}

For three integers $0\leq f, g, h<2^n-1$ satisfying $h\equiv f+g\mbox{ mod}(2^n-1)$, let $f=\sum_{i=0}^{n-1}f_i2^{i}$, $g=\sum_{i=0}^{n-1}g_i2^{i}$,
and $h=\sum_{i=0}^{n-1}h_i2^{i}$, where $f_i, g_i, h_i\in\{0, 1\}$ for any $0\leq i<n$. Then there exists a unique integer sequence
$\overrightarrow{c}=c_0, c_1, ..., c_{n-1}$ with $c_i\in\{0, 1\}$ for any $0\leq i<n$ satisfying
$$h_i+2c_i=f_i+g_i+c_{i-1},\ 0\leq i\leq n-1,$$
where $c_{i-1}=c_{n-1}$ if $i=0$. Let $\mathrm{wt}(\overrightarrow{c})=c_0+c_1+...+c_{n-1}$. Then we have
$$\mathrm{wt}(f)+\mathrm{wt}(g)=\mathrm{wt}(h)+\mathrm{wt}(\overrightarrow{c})\geq\mathrm{wt}(f+g).$$

The following lemma is known.
\begin{lemma}[\cite{LL08}]\label{lem_k}
For any $0<j\leq2^k$, we have $\mathrm{wt}(j(2^k-1))=k$.
\end{lemma}

%===========================================================================
%===========================================================================
\section{The Condition of Langevin and Leander}\label{sec_main_1}

Suppose that $f(x)=Tr_1^n(\alpha x^d)$ is a bent function, where $0<d<2^n-1$ and $\alpha\in \mathbb{F}_{2^n}^{*}$. Let
$$\mathcal{J}_d=\{\ \widetilde{j}\ |\ 0<\widetilde{j}<2^n-1, V_d(\widetilde{j})=\min_{0<j<2^n-1}V_d(j)\ \},$$
and
$$\Pi_d(x)=\sum_{j\in \mathcal{J}_d}x^j\in \mathbb{F}_2[x].$$

If $\min_{0<j<2^n-1}V_d(j)=k$, we have the following theorem. Most part of this proof is the same as that of Theorem 4 in \cite{LL08}. For completeness, we provide the whole proof.
\begin{theorem}\label{thm_=k}
With notation as above, if $\min_{0<j<2^n-1}V_d(j)=k$, and $V_d(\widetilde{j})=k$ with $0<\widetilde{j}<2^n-1$, then $\widetilde{j}d\equiv0\mod(2^n-1)$.
Moreover, we have $\Pi_d(\alpha)=1$.
\end{theorem}
{\bf Proof. }By Lemma \ref{lem_gauss_trace}, if $\lambda\neq0$, we have
\begin{eqnarray*}
\widehat{f}(\lambda)&=&\sum_{x\in\mathbb{F}_{2^n}}(-1)^{Tr_1^n(\alpha x^d)+Tr_1^n(\lambda x)}\\
&=&1+\frac{1}{(2^n-1)^2}\sum_{x\in\mathbb{F}_{2^n}^{*}}\sum_{\chi_1\in\widehat{\mathbb{F}_{2^n}^{*}}}G(\chi_1)\overline{\chi_1}(\alpha x^d)\sum_{\chi_2\in\widehat{\mathbb{F}_{2^n}^{*}}}G(\chi_2)\overline{\chi_2}(\lambda x)\\
&=&1+\frac{1}{(2^n-1)^2}\sum_{\chi_1,\chi_2\in\widehat{\mathbb{F}_{2^n}^{*}}}G(\chi_1)G(\chi_2)\overline{\chi_1}(\alpha)\overline{\chi_2}(\lambda)\sum_{x\in\mathbb{F}_{2^n}^{*}}\overline{\chi_1}(x^d)\overline{\chi_2}(x)\\
&=&1+\frac{1}{2^n-1}\sum_{\chi\in\widehat{\mathbb{F}_{2^n}^{*}}}G(\chi)G(\overline{\chi}^d)\overline{\chi}(\alpha)\chi^d(\lambda).
\end{eqnarray*}
For $\chi\in\widehat{\mathbb{F}_{2^n}^{*}}$, let $\chi=\chi_\mathfrak{p}^{-j}$ with $1\leq j\leq 2^n-2$, we get
\begin{eqnarray*}
\widehat{f}(\lambda)&=&1+\frac{1}{2^n-1}+\frac{1}{2^n-1}\sum_{j=1}^{2^n-2}G(\chi_\mathfrak{p}^{-j})G(\chi_\mathfrak{p}^{jd})\chi_\mathfrak{p}^{j}(\alpha)\chi_\mathfrak{p}^{-jd}(\lambda)\\
&=&\frac{2^n}{2^n-1}+\frac{1}{2^n-1}\sum_{j=1}^{2^n-2}G(\chi_\mathfrak{p}^{-j})G(\chi_\mathfrak{p}^{jd})\chi_\mathfrak{p}^{j}(\alpha)\chi_\mathfrak{p}^{-jd}(\lambda).
\end{eqnarray*}
Because $f(x)$ is a bent function, it follows that
\begin{eqnarray*}
\frac{2^k}{2^n-1}+\frac{\sum_{j=1}^{2^n-2}G(\chi_\mathfrak{p}^{-j})G(\chi_\mathfrak{p}^{jd})\chi_\mathfrak{p}^{j}(\alpha)\chi_\mathfrak{p}^{-jd}(\lambda)}{(2^n-1)2^k}&=&\frac{\widehat{f}(\lambda)}{2^k}=\pm1.
\end{eqnarray*}
Therefore, we have
\begin{eqnarray*}
\lefteqn{\left(\frac{2^k}{2^n-1}+\frac{\sum_{j=1}^{2^n-2}G(\chi_\mathfrak{p}^{-j})G(\chi_\mathfrak{p}^{jd})\chi_\mathfrak{p}^{j}(\alpha)\chi_\mathfrak{p}^{-jd}(\lambda)}{(2^n-1)2^k}\right)(\mbox{mod }\mathcal{P})}\\
&=&\frac{\sum_{j=1}^{2^n-2}G(\chi_\mathfrak{p}^{-j})G(\chi_\mathfrak{p}^{jd})\chi_\mathfrak{p}^{j}(\alpha)\chi_\mathfrak{p}^{-jd}(\lambda)}{2^k}(\mbox{mod }\mathcal{P})\\
&=&\sum_{j\in \mathcal{J}_d}G(\chi_\mathfrak{p}^{-j})G(\chi_\mathfrak{p}^{jd})\chi_\mathfrak{p}^{j}(\alpha)\chi_\mathfrak{p}^{-jd}(\lambda)(\mbox{mod }\mathcal{P})\\
&=&\sum_{j\in \mathcal{J}_d}\alpha^j\lambda^{-jd}=1.
\end{eqnarray*}
Let $g_\alpha(\lambda)=\sum_{j\in \mathcal{J}_d}\alpha^j\lambda^{-jd}+1$. Then, for any fixed $\alpha$, $g_\alpha(\lambda)$ is a polynomial in $\lambda$ with degree at most $2^n-2$. However, there are $2^n-1$ solutions of $g_\alpha(\lambda)=0$. Therefore, $g_\alpha(\lambda)$ is a constant polynomial, and $jd\equiv0\mod(2^n-1)$ for any $j\in \mathcal{J}_d$. Thus, $\Pi_d(\alpha)=1$. \done

On the other hand, if $\min_{0<j<2^n-1}V_d(j)<k$, we have the following similar result.

\begin{theorem}\label{thm_<k}
With notation as above, if $\min_{0<j<2^n-1}V_d(j)<k$, and $V_d(\widetilde{j})=\min_{0<j<2^n-1}V_d(j)$ with $0<\widetilde{j}<2^n-1$, then $\widetilde{j}d\equiv0\mod(2^n-1)$. Moreover, we have $\Pi_d(\alpha)=0$.
\end{theorem}
{\bf Proof. }Let $t=\min_{0<j<2^n-1}V_d(j)$. Similar to the proof of Theorem \ref{thm_=k}, if $\lambda\neq0$, we have
\begin{eqnarray*}
\widehat{f}(\lambda)=\frac{2^n}{2^n-1}+\frac{1}{2^n-1}\sum_{j=1}^{2^n-2}G(\chi_\mathfrak{p}^{-j})G(\chi_\mathfrak{p}^{jd})\chi_\mathfrak{p}^{j}(\alpha)\chi_\mathfrak{p}^{-jd}(\lambda).
\end{eqnarray*}
Because $f(x)$ is a bent function, it follows that
\begin{eqnarray*}
\frac{2^{n-t}}{2^n-1}+\frac{\sum_{j=1}^{2^n-2}G(\chi_\mathfrak{p}^{-j})G(\chi_\mathfrak{p}^{jd})\chi_\mathfrak{p}^{j}(\alpha)\chi_\mathfrak{p}^{-jd}(\lambda)}{(2^n-1)2^t}&=&\frac{\widehat{f}(\lambda)}{2^t}=\pm2^{k-t}.
\end{eqnarray*}
Therefore, we have
\begin{eqnarray*}
\left(\frac{2^{n-t}}{2^n-1}+\frac{\sum_{j=1}^{2^n-2}G(\chi_\mathfrak{p}^{-j})G(\chi_\mathfrak{p}^{jd})\chi_\mathfrak{p}^{j}(\alpha)\chi_\mathfrak{p}^{-jd}(\lambda)}{(2^n-1)2^t}\right)(\mbox{mod }\mathcal{P})=\sum_{j\in \mathcal{J}_d}\alpha^j\lambda^{-jd}=0.
\end{eqnarray*}
The left part of the proof is the same as that of Theorem \ref{thm_=k}, so we omit the details. \done

In the following, we study the case of Dillon exponent in more detail.

\begin{lemma}\label{lem_dillon}
Let $k\geq 3$, and $d=2^k-1$. Then it holds that $\min_{0<j<2^n-1}V_d(j)=2$. Moreover, we have $\mathcal{J}_d=\{2^k+1, 2(2^k+1), 2^2(2^k+1), ..., 2^{k-1}(2^k+1)\}$.
\end{lemma}
{\bf Proof. }Let $t=\min_{0<j<2^n-1}V_d(j)$. Then $t\leq V_d(2^k+1)=2<k$. By Theorem \ref{thm_<k}, for any $0<j<2^n-1$, if $V_d(j)=t$, then $jd\equiv0\mod(2^n-1)$. Therefore, we get $j=(2^k+1)\widetilde{j}$, where $0<\widetilde{j}<2^k-1$. If $\widetilde{j}\not\in\{1, 2, 2^2, ..., 2^{k-1}\}$, then $V_d(j)=\mathrm{wt}((2^k+1)\widetilde{j})>2$. Thus, it follows that $t=2$ and $\mathcal{J}_d=\{2^k+1, 2(2^k+1), 2^2(2^k+1), ..., 2^{k-1}(2^k+1)\}$.\done

If $\alpha\in \mathbb{F}_{2^k}^{*}$, and $d=2^k-1$. Then $f(x)=Tr_1^n(\alpha x^d)$ is a bent function if $\sum_{x\in\mathbb{F}_{2^k}^{*}}(-1)^{Tr_1^k(\alpha x+1/x)}=-1$. By Lemma \ref{lem_dillon}, $\mathcal{J}_d=\{2^k+1, 2(2^k+1), 2^2(2^k+1), ..., 2^{k-1}(2^k+1)\}$. By Theorem \ref{thm_<k}, $\Pi_d(\alpha)=0$. Let us compute
$$\Pi_d(\alpha)=\sum_{j\in \mathcal{J}_d}\alpha^j=\sum_{i=0}^{k-1}\alpha^{2^i(2^k+1)}=\sum_{i=0}^{k-1}\alpha^{2^{i+1}}=Tr_1^k(\alpha).$$
Therefore, $Tr_1^k(\alpha)=0$. As a consequence, we get the following proposition.
\begin{proposition}\label{prop_Kloosterman}
Let $k\geq 3$. For any $\alpha\in \mathbb{F}_{2^k}$, if $\sum_{x\in\mathbb{F}_{2^k}^{*}}(-1)^{Tr_1^k(\alpha x+1/x)}=-1$, then $Tr_1^k(\alpha)=0$.
\end{proposition}

\begin{remark}
The result of Proposition \ref{prop_Kloosterman} is already covered in \cite{CHZ09} and \cite{HZ99}, which confirms the correctness of Theorem \ref{thm_<k}.
Following the approach in this paper, it is possible to get more nice properties about Kloosterman sums with the help of more information about $k$ and better tools such as the Gross-Koblitz formula \cite{Koblitz80}.
\end{remark}

\begin{remark}
In 2008, if $k$ is even and $k>4$, Charpin and Gong showed that $\sum_{x\in\mathbb{F}_{2^k}^{*}}(-1)^{Tr_1^k(\alpha x+1/x)}\neq-1$ for any $\alpha\in \mathbb{F}_{2^{k/2}}^{*}$ \cite{CG08}.
In 2009, Shparlinski generalized this result \cite{Shparlinski09}, and Moisio finished the proof of the subfield conjecture \cite{Moisio09}.
\end{remark}

%===========================================================================
%===========================================================================
\section{The Open Problem of Pott {\em et al.}}\label{sec_main_2}

Let $F(x)=x^d$, where $0<d<2^n-1$. In this section, we dig out more properties about $\mathcal{S}_F$ and $d$, and solve Open Problem \ref{open}.

In 2006, Leander proved the following nice lemma.

\begin{lemma}[\cite{Leander06}]\label{lem_leander}
With notation as above, let $t=\gcd(2^n-1, d)$. If $f(x)=Tr_1^n(\alpha x^d)$ is a bent function for some $\alpha\in\mathbb{F}_{2^n}$, then there are two cases of $t$:
1) $t|(2^k+1)$; 2) $t|(2^k-1)$.
\end{lemma}

\begin{lemma}\label{lem_square}
With notation as above, if $\alpha\in \mathcal{S}_F$, then $\alpha^2\in \mathcal{S}_F$.
\end{lemma}
{\bf Proof. }Let $f_\alpha(x)=Tr_1^n(\alpha x^d)$, and $f_{\alpha^2}(x)=Tr_1^n(\alpha^2 x^d)$. Then, for any $\lambda$, we have
$$\widehat{f}_{\alpha^2}(\lambda)=\sum_{x\in\mathbb{F}_{2^n}}(-1)^{Tr_1^n(\alpha^2 x^d+\lambda x)}=\sum_{x\in\mathbb{F}_{2^n}}(-1)^{Tr_1^n(\alpha^2 x^{2d}+\lambda x^2)}=\sum_{x\in\mathbb{F}_{2^n}}(-1)^{Tr_1^n(\alpha x^d+\sqrt{\lambda} x)}=\widehat{f}_{\alpha}(\sqrt{\lambda}).$$
Thus, if $\alpha\in \mathcal{S}_F$, then $\alpha^2\in \mathcal{S}_F$. \done

For any $\alpha\in \mathbb{F}_{2^n}$, let $O(\alpha)$ be the smallest integer $t>0$ satisfying $\alpha^{2^t}=\alpha$. Then $\alpha\in \mathbb{F}_{2^{O(\alpha)}}$. Moreover, let $\mathcal{L}(\alpha)$ be the linear space over $\mathbb{F}_2$ generated by $\alpha, \alpha^2, ..., \alpha^{2^{O(\alpha)}}$. Then $\mathcal{L}(\alpha)=\mathbb{F}_{2^{O(\alpha)}}$. In particular, if $O(\alpha)=n$, then $\mathcal{L}(\alpha)=\mathbb{F}_{2^n}$.

\begin{lemma}\label{lem_k_even}
Suppose that $k=2^ep_1^{e_1}p_2^{e_2}...p_t^{e_t}$, where $t\geq1, e\geq1, e_i\geq1$, and $p_i$ are prime numbers satisfying $3\leq p_1<p_2<...<p_t$. Then we have
$$2^{2k/p_1}+2^{2k/p_2}+...+2^{2k/p_t}+2^{k/2}<2^k.$$
\end{lemma}

{\bf Proof. }There are two cases.

1) $p_1=3$. In this case, we have $2^{2k/p_1}+2^{2k/p_2}+...+2^{2k/p_t}+2^{k/2}<(t+1)2^{2k/3}$. Thus, we only need to prove that $(t+1)2^{2k/3}\leq 2^k$, which is equivalent to $t+1\leq 2^{k/3}$. One may check that $2^{k/3}\geq 2^{3^{t-1}}\geq t+1$. The result follows.

2) $p_1>3$. In this case, we have $2^{2k/p_1}+2^{2k/p_2}+...+2^{2k/p_t}+2^{k/2}<(t+1)2^{k/2}$. Thus, we only need to prove that $(t+1)2^{k/2}\leq 2^k$, which is equivalent to $t+1\leq 2^{k/2}$. One may check that $2^{k/2}>2^{k/3}\geq t+1$. The result follows, too.\done

\begin{lemma}\label{lem_k_odd}
Suppose that $k=p_1^{e_1}p_2^{e_2}...p_t^{e_t}$, where $t\geq1, e_i\geq1$, and $p_i$ are prime numbers satisfying $3\leq p_1<p_2<...<p_t$. The we have
$$2^{2k/p_1}+2^{2k/p_2}+...+2^{2k/p_t}<2^k.$$
\end{lemma}

{\bf Proof. }The proof is similar to that of Lemma \ref{lem_k_even}, so we omit the details.\done

\begin{lemma}\label{lem_subfield}
With notation as above, if $|\mathcal{S}_F|=2^k$, then $\mathcal{S}_F=\mathbb{F}_{2^k}$.
\end{lemma}
{\bf Proof. }Suppose that $\mathcal{S}_F\not=\mathbb{F}_{2^k}$. Let $\mathcal{S}_1=\mathcal{S}_F\setminus\mathbb{F}_{2^k}$, and $\mathcal{S}_2=\mathcal{S}_F\bigcap\mathbb{F}_{2^k}$. Then $|\mathcal{S}_1|\geq1$, and $|\mathcal{S}_2|<2^k$. For simplicity, let $\alpha\in\mathcal{S}_1$, and $\beta\in\mathcal{S}_2$. By Theorem \ref{thm_S} and Lemma \ref{lem_square}, we have $\mathcal{L}(\alpha),\mathcal{L}(\beta)\subseteq\mathcal{S}_F$.
If $O(\alpha)=n$, then $\mathbb{F}_{2^n}\subseteq\mathcal{S}_F$. We get a contradiction. If $O(\beta)=k$, then $\mathbb{F}_{2^k}\subseteq\mathcal{S}_F$. We get a contradiction again. Hence, in the following, we can assume that $O(\alpha)<n$, and $O(\beta)<k$.

We divide the proof into three cases.

1) $k$ is a prime number. There are two subcases.

(1) $k=2$. In this subcase, $O(\alpha)|k$. Thus, $\alpha\in\mathbb{F}_{2^k}$, which is a contradiction.

(2) $k>2$. In this subcase, $O(\alpha)|2$, and $O(\beta)=1$. Thus, $\alpha\in\mathbb{F}_{2^2}$, and $\beta\in\mathbb{F}_{2}$. It follows that $\mathcal{S}_F=\mathcal{S}_1\cup\mathcal{S}_2\subseteq\mathbb{F}_{2^2}$.
Therefore, $|\mathcal{S}_F|\leq2^2<2^k$. This is a contradiction.

2) $k$ is an even composite number. There are three subcases.

(1) $k=2^e$, where $e\geq2$. In this subcase, $O(\alpha)|k$. Thus, $\alpha\in\mathbb{F}_{2^k}$, which is a contradiction.

(2) $k=2^ep_1^{e_1}$, where $e\geq1, e_1\geq1$, and $p_1\geq3$ is prime. In this subcase, $O(\alpha)|2^{e+1}p_1^{e_1-1}$, and $O(\beta)|2^{e-1}p_1^{e_1}$ or $O(\beta)|2^ep_1^{e_1-1}$.
Thus, $\alpha\in\mathbb{F}_{2^{2k/p_1}}$, and $\beta\in\mathbb{F}_{2^{k/2}}$ or $\beta\in\mathbb{F}_{2^{k/p_1}}$. It follows that $\mathcal{S}_F=\mathcal{S}_1\cup\mathcal{S}_2\subseteq\mathbb{F}_{2^{2k/p_1}}\cup\mathbb{F}_{2^{k/2}}$. Therefore, by Lemma \ref{lem_k_even}, we get $$|\mathcal{S}_F|\leq2^{2k/p_1}+2^{k/2}<2^k.$$ This is a contradiction.

(3) $k=2^ep_1^{e_1}p_2^{e_2}...p_t^{e_t}$, where $t\geq2, e\geq1, e_i\geq1$, and $p_i$ are prime numbers satisfying $3\leq p_1<p_2<...<p_t$.
Similarly, in this subcase, we have $\alpha\in\mathbb{F}_{2^{2k/p_1}}\cup\mathbb{F}_{2^{2k/p_2}}\cup...\cup\mathbb{F}_{2^{2k/p_t}}$, and $\beta\in\mathbb{F}_{2^{k/2}}\cup\mathbb{F}_{2^{k/p_1}}\cup...\cup\mathbb{F}_{2^{k/p_t}}$.
It follows that $$\mathcal{S}_F=\mathcal{S}_1\cup\mathcal{S}_2\subseteq\mathbb{F}_{2^{2k/p_1}}\cup\mathbb{F}_{2^{2k/p_2}}\cup...\cup\mathbb{F}_{2^{2k/p_t}}\cup\mathbb{F}_{2^{k/2}}.$$ Therefore, by Lemma \ref{lem_k_even}, we get
$$|\mathcal{S}_F|\leq2^{2k/p_1}+2^{2k/p_2}+...+2^{2k/p_t}+2^{k/2}<2^k.$$ This is a contradiction.

3) $k$ is an odd composite number. There are two subcases.

(1) $k=p_1^{e_1}$, where $e_1\geq2$, and $p_1\geq3$ is prime. In this subcase, we have $\alpha\in\mathbb{F}_{2^{2k/p_1}}$, and $\beta\in\mathbb{F}_{2^{k/p_1}}$. It follows that $\mathcal{S}_F=\mathcal{S}_1\cup\mathcal{S}_2\subseteq\mathbb{F}_{2^{2k/p_1}}$. Therefore, $|\mathcal{S}_F|\leq2^{2k/p_1}<2^k$. This is a contradiction.

(2) $k=p_1^{e_1}p_2^{e_2}...p_t^{e_t}$, where $t\geq2, e_i\geq1$, and $p_i$ are prime numbers satisfying $3\leq p_1<p_2<...<p_t$. In this subcase, we have $\alpha\in\mathbb{F}_{2^{2k/p_1}}\cup\mathbb{F}_{2^{2k/p_2}}\cup...\cup\mathbb{F}_{2^{2k/p_t}}$, and $\beta\in\mathbb{F}_{2^{k/p_1}}\cup...\cup\mathbb{F}_{2^{k/p_t}}$.
It follows that $$\mathcal{S}_F=\mathcal{S}_1\cup\mathcal{S}_2\subseteq\mathbb{F}_{2^{2k/p_1}}\cup\mathbb{F}_{2^{2k/p_2}}\cup...\cup\mathbb{F}_{2^{2k/p_t}}.$$
Therefore, by Lemma \ref{lem_k_odd}, we get
$$|\mathcal{S}_F|\leq2^{2k/p_1}+2^{2k/p_2}+...+2^{2k/p_t}<2^k.$$ This is a contradiction, too.\done

\begin{remark}
For the case of binomial or multinomial functions over $\mathbb{F}_{2^{2k}}$, Lemma \ref{lem_subfield} still holds.
\end{remark}

\begin{lemma}\label{lem_divisor}
With notation as above, if $|\mathcal{S}_F|=2^k$, then $d=(2^k+1)s$ with $0<s<2^k-1$ and $\gcd(s, 2^k-1)=1$.
\end{lemma}
{\bf Proof. }Suppose that $(2^k+1)\not|d$. Let $\alpha$ be a primitive element of $\mathbb{F}_{2^n}$. Then $\alpha^d\not\in\mathbb{F}_{2^k}$. By Lemma \ref{lem_subfield}, $1\in\mathcal{S}_F$.
It follows that $\alpha^d\in \mathcal{S}_F$. However, by Lemma \ref{lem_subfield}, we know $\mathcal{S}_F=\mathbb{F}_{2^k}$. Hence, $\alpha^d\not\in \mathcal{S}_F$. This is a contradiction. Hence, we get $(2^k+1)|d$.
Furthermore, by Lemma \ref{lem_leander}, we know $d=(2^k+1)s$ with $0<s<2^k-1$ and $\gcd(s, 2^k-1)=1$.\done

\begin{theorem}\label{thm_coprime}
Let $d=(2^k+1)s$ with $0<s<2^k-1$ and $\gcd(s, 2^k-1)=1$. If $s\not\in\{1, 2, 2^2, ..., 2^{k-1}\}$, then $f(x)=Tr_1^n(\alpha x^d)$ is not a bent function for any $\alpha\in\mathbb{F}_{2^n}$.
\end{theorem}
{\bf Proof. }Suppose that there exists $\alpha\in\mathbb{F}_{2^n}$ such that $f(x)=Tr_1^n(\alpha x^d)$ is a bent function. Let $0<t<2^k-1$ and $t\equiv s^{-1}(\mbox{mod }2^k-1)$. Because $s\not\in\{1, 2, 2^2, ..., 2^{k-1}\}$, we have $2\leq\mathrm{wt}(t)<k$. Let $\widetilde{t}=2^k-1-t$. Then $0<\mathrm{wt}(\widetilde{t})\leq k-2$.
Consequently, we compute $$V_d(\widetilde{t})=\mathrm{wt}(\widetilde{t})+\mathrm{wt}(-\widetilde{t}d)=\mathrm{wt}(\widetilde{t})+\mathrm{wt}((2^k+1)st)=\mathrm{wt}(\widetilde{t})+\mathrm{wt}(2^k+1)\leq k.$$

There are two cases.

1) $\min_{0<j<2^n-1}V_d(j)=k$. In this case, $V_d(\widetilde{t})=k$. By Theorem \ref{thm_=k}, if $j\in \mathcal{J}_d$, then  $jd\equiv0\mod(2^n-1)$. Therefore, we have $(2^k-1)|j$. However, $(2^k-1)\not|\widetilde{t}$. This is a contradiction.

2) $\min_{0<j<2^n-1}V_d(j)<k$. Similarly, by Theorem \ref{thm_<k}, if $j\in \mathcal{J}_d$, then we have $(2^k-1)|j$. However, by Lemma \ref{lem_k}, we get $\mathrm{wt}(j)=k$. This is a contradiction, too.

Hence, for any $\alpha\in\mathbb{F}_{2^n}$, $f(x)=Tr_1^n(\alpha x^d)$ is not a bent function.\done

\begin{theorem}
$x^{2^k+1}$ is the only monomial function over $\mathbb{F}_{2^{2k}}$ which has maximal number of bent components.
\end{theorem}
{\bf Proof. }By Lemma \ref{lem_divisor} and Theorem \ref{thm_coprime}, the result follows. \done

%===========================================================================
%===========================================================================
\section{Conclusion}\label{sec_con}

In this paper, we solve an open problem of Pott {\em et al.} regarding bent components in 2018, and show that  $x^{2^k+1}$ is the only monomial function over $\mathbb{F}_{2^{2k}}$ with maximal number of bent components. We find that this open problem is closely related to the strong condition of Langevin and Leander in 2008. Via new observations, we dig out interesting and pretty useful properties about this condition. Following the approach in this paper, it is possible to get more nice properties about Kloosterman sums with the help of more information about $k$ and better tools from number theory such as the Gross-Koblitz formula \cite{Koblitz80}. Moreover, we also solve an open problem of Ness and Helleseth in 2006 because it is essentially the same as the open problem of Pott {\em et al.} in 2018.

%===========================================================================
%===========================================================================
\section*{Acknowledgement}

The authors would like to thank Prof. Pascale Charpin for her invaluable comments on Kloosterman sums, and Dr. Lijing Zheng for his nice suggestions.

%===========================================================================
%===========================================================================

\end{document}